\documentclass[12pt]{article}

\usepackage{latexsym}
\usepackage{graphicx}
\usepackage{color}
\usepackage{multicol}

\begin{document}


\begin{center}
{\large \textbf{``Physics in Films''}}\\
{\large \textbf{A New Approach to Teaching Science}}\\[5mm]

 C. Efthimiou\footnote{costas@physics.ucf.edu}
    and
 R. Llewellyn\footnote{ral@physics.ucf.edu}\\
 Physics Department\\
 University of Central Florida\\
 Orlando, FL 32816
\end{center}

{\small
\begin{abstract}
Over the past year and a half we have developed an innovative
approach to the teaching of \textit{Physical Science}, a general
education course typically found in the curricula of nearly every
college and university. The new approach uses popular movies to
illustrate the principles of physical science, analyzing
individual scenes against the background of the fundamental
physical laws. The impact of being able to understand
\textit{why}, in reality, the scene could or could not have
occurred as depicted in the film, what the director got right and
what he got wrong, has excited student interest enormously in a
course that, when taught in the traditional mode, is usually
considered to be `too hard and boring'. The performance of
students on exams reflected the increased attention to and
retention of basic physical concepts, a result that was a primary
goal of the `\textit{Physics in Films}' approach. Following the
first offering of the revitalization of the Physical Science
course, in which action and sci-fi films were the primary source
of the scene clips used in class, the instructors have
demonstrated the versatility of the approach by building
variations of the course around other genres, as well
---\textit{Physics in Films: Superheroes} and \textit{Physics in
Films: Pseudoscience}. A parallel approach to the general
education course in astronomy is currently being discussed; many
others are in our thoughts.
\end{abstract}


}


\vspace{5mm}


\begin{center}
\textbf{BACKGROUND}
\end{center}

It is well documented that interest and understanding of science
among people of all ages in the United States has declined
severely and currently stands at an alarmingly low level.
According to surveys conducted by the National Science Foundation
(NSF 2002) over many years, about 50 percent of the people do not
know that Earth takes one year to go around the sun, that
electrons are smaller than atoms, and that early humans did not
live at the same time as the dinosaurs. These examples of faulty
knowledge of physics surely extend to other sciences and are
mirrored in other nations. This trend is due in part to a changing
society that encourages new generations to adopt a more
materialistic ideology and a disrespect, even distrust of
scientific knowledge. Even though the luxury of technology we
currently enjoy is due in large measure to science, our society
has chosen to forget that, tacitly assuming that such progress is
a natural phenomenon that occurs automatically and about which one
need not worry.


\begin{figure}[h!]
\begin{center}
\includegraphics[width=3in]{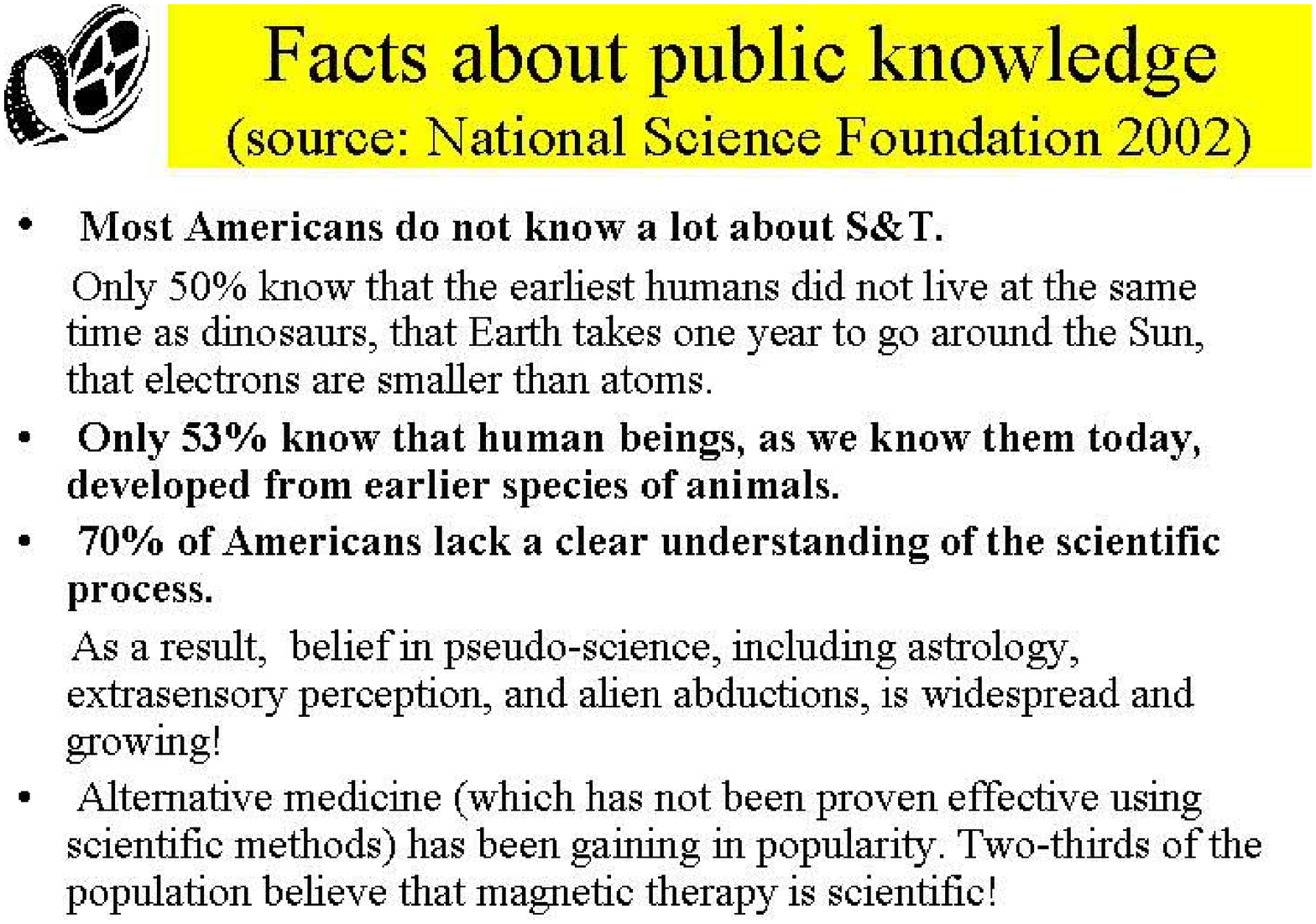}
\label{fig1}
\end{center}
\caption{Some facts about science literacy in USA.}
\end{figure}


\begin{center}
{\bfseries THE CORE IDEA AND THE FIRST IMPLEMENTATION}
\end{center}

The project addresses the issue of how to instill in the broad spectrum of
college and university students the enthusiasm and excitement of physics all
physicists have experienced and continue to experience. We have thus
proposed to accomplish this by adopting as a teaching vehicle a medium that
the students have already accepted as a reflection of today's culture,
namely by using popular movies to illustrate both the basic principles and
frontier discoveries of science (Efthimiou {\&} Llewellyn 2003). The
targeted audience is science and non-science majors alike. The course we
have created is more relevant and, frankly, a more interesting substitute
for the traditional Physical Science courses taught in nearly all colleges
and universities. If this effort proves to be as successful as our early
results suggest that it will be, an appropriate version for majors in other
science disciplines is a definite option for the future.

\begin{center}
\textbf{Summer 2002: Action and Sci-Fi Movies!}
\end{center}

During the initial offering of the course in Summer 2002 we discussed the
principles of physics using scene clips from popular action and sci-fi
movies. For example, the law of gravitation as (mis)used in \textit{Independence Day},
conservation
of momentum in \textit{Tango and Cash}, speed and acceleration in \textit{Speed 2}, and so on.
Figure 2 shows the
nine movies used that first summer. Students were required to watch the
films at home and turn in a brief, written analysis of the physics principle
illustrated in each of three scenes of their own choosing (homework!). In
class, five to ten percent of the class (of 90) were called upon each day to
orally present their analysis of one scene to the class. Both the written
and oral analyses became part of their grade in the course.


\begin{figure}[h!]
\begin{center}
\includegraphics[width=3in]{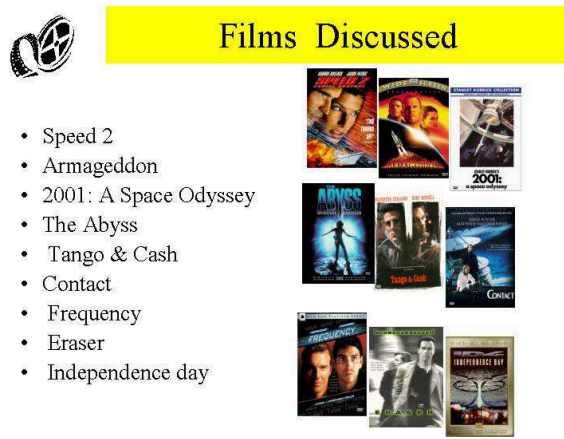}
\end{center}
\caption{Action and Sci-Fi films used in the original approach of
        the \textit{Physics In Films} variation of
        the \textit{Physical Science} course.}
\label{fig2}
\end{figure}


Based on the data collected so far and on discussions among the
several physical science instructors and hundreds of physical
science students at our institution (physical science enrollment:
2600 students per year), we have had an overwhelming success (APS
News 2003, Chow 2003, Graham 2002, Grayson 2002, Priore 2003). Our
experiences thus far reveal a strong tendency by the students to
participate enthusiastically in discussions of physical science,
if the topic under discussion is familiar from TV shows or movies.
Not only may such a course be widely emulated and serve to educate
our society, but it may also help correct misconceptions of
science that popular movies and TV series have created,
misconceptions that we feel have contributed to the public
mistrust of science.


\begin{figure}[h!]
\begin{center}
\includegraphics[width=3in]{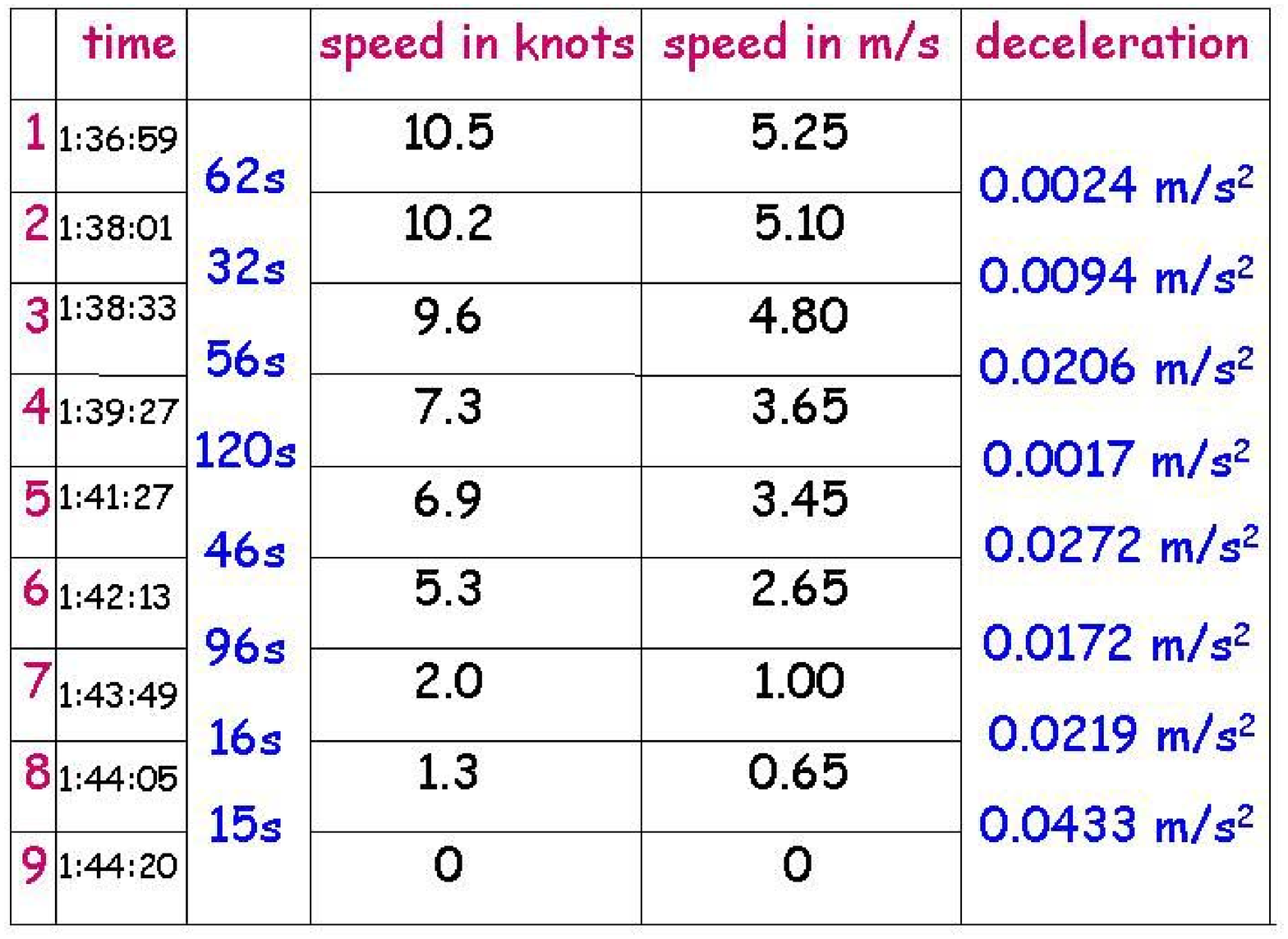}
\end{center}
\caption{The deceleration of the cruise ship according to the
         director's data in \textit{Speed 2} as it crashes
         on the port.}
\label{fig3}
\end{figure}


\begin{center}
\textbf{Example---}\textbf{\textit{Speed 2}}
\end{center}

In \textit{Speed 2} (starring Sandra Bullock) there is a scene where a large cruise ship
crashes into the waterfront of a busy resort town, hitting other boats,
docks, the shoreline, and ultimately buildings. As the ship approaches the
shore and hits various things, it begins to decelerate and people and
objects onboard are thrown violently toward the bow, two men on the bridge
even thrown through the windows onto the forward deck. All of the mayhem is
seemingly in agreement with the audience's `gut' feeling of what would
happen as a consequence of the deceleration of the ship. However, throughout
the scene we are given frequent views of the ship's digital speedometer on
the bridge. Since the crashing of the ship is shown in real time, by simply
noting the time of each speed reading as the ship approaches the shore and
finally comes to a stop, we can directly compute the deceleration, using the
simple formula:
$$
    deceleration\,=\,\frac{speed\,change}{time\,change} ~.
$$
The table in figure 3 shows the results of such a calculation
based on the director's data. The numbers in the second column are
clock times (according to the DVD timer) at which speeds (in
knots) were read from the ship's meter. The numbers in the third
column (lined up with the horizontal lines) are the seconds
between successive readings. The fourth column converts the speeds
into meters per second (1 knot  is about a half meter per second)
and the right-hand column shows the corresponding decelerations in
meters per second square (m/s$^{2})$. Comparing the decelerations
in the last column with the acceleration of gravity, about 10
m/s$^{2}$, we see that the ship's deceleration was actually quite
low. To understand how low, think about the deceleration you would
experience in bringing your car, travelling at 30 miles per hour,
to a gentle stop (that is, applying the brakes for about 10
seconds) at a red traffic signal; this acceleration would be about
1.33 m/s$^{2}$, or more than 30 times the greatest deceleration
0.0433 m/s$^{2 }$of the table. At this point the students are
amazed! The ship certainly has an enormous amount of kinetic
energy and that energy is dissipated in the collisions, but the
widespread tumbling about of things and people on the ship would
not have occurred. Indeed, a passenger who happened to be asleep
in a bunk might not have even noticed the collision. The concept
of impulse can also be discussed with the aid of this scene. The
forces that people and objects on the ship experience during the
collision are quite low because (1) the ship is not going very
fast to begin with (hence, momenta are not large) and (2) the
collision takes place over a very long time, more than 10 minutes.

\begin{figure}[h!]
\begin{center}
\includegraphics[width=3in]{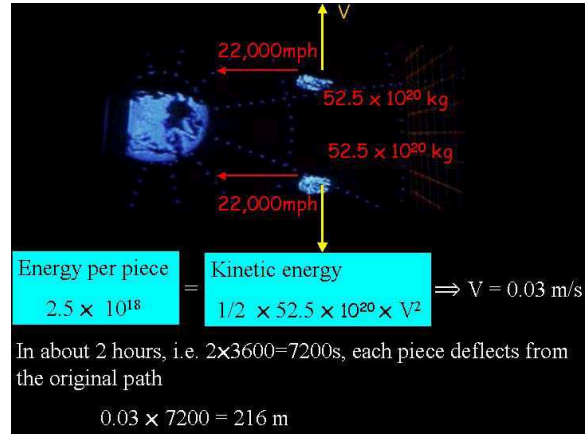}
\end{center}
\caption{Summary of the calculation for the deflection of the
         asteroid fragments in \textit{Armageddon}.}
\label{fig4}
\end{figure}


\begin{center}
\textbf{Example---}\textbf{\textit{Armageddon}}
\end{center}

In \textit{Armageddon} (starring Bruce Willis) a huge, errant asteroid the size of Texas is on
a collision course with Earth. A team of oil well drillers is dispatched via
a pair of space shuttles to intercept the asteroid, drill a hole in it at
the right place, lower a large nuclear bomb into the hole, and subsequently
blow the asteroid into two large pieces. The transverse velocities imparted
to the two pieces by the explosion, when added to their (undiminished)
velocities toward Earth, are to cause the pieces to just miss Earth, thereby
averting worldwide disaster. Analysis of this situation uses conservation of
energy, conservation of momentum, vector addition, and the law of gravity. A
summary of the overall situation is depicted in the diagram of figure 4.

Using numbers provided in the film, we introduce the students to
the idea of making reasonable approximations. For example, the
asteroid is, we are told, the size of Texas, so we assume Texas is
a square whose surface area equals that of the state, then
approximate the asteroid as a cube, each of whose sides equals the
surface area of the state. Multiplying the volume of the cube by
the average density of Earth gives us a decent estimate of the
mass of the asteroid. Assuming the bomb as being equal to 100,000
Hiroshima bombs provided an estimate of the energy available for
the job. Then, assuming \textit{all} of that energy became kinetic
energy equally divided between the two pieces of the asteroid
(i.e., ignoring the energy needed to break the asteroid into two
pieces), we can readily compute the distance the pieces have moved
perpendicular to their original direction of motion by the time
they reach Earth. As noted in the diagram, the deflection for each
piece is only a bit over 200 meters. Once again the students are
astonished. Instead of being hit by one Texas-size asteroid, Earth
will be hit by two half-Texas-size asteroids few city-blocks
apart! We then wind up this discussion with an explanation of what
is realistically possible and why the government has an ongoing
project to detect and track space objects approaching Earth or in
Earth-crossing orbits.

\begin{center}
\textbf{BEYOND ACTION AND SCI-FI MOVIES}
\end{center}

After the pilot course was very successfully tested in four
sections (with total enrollment of 800 students), the authors were
motivated to develop the course further and to explore new
directions. The original pilot course included movies that were
selected to span the entire topical range of the standard
\textit{Physical Science} course. In the selection of the movies
no attention was paid to the genre or the theme of the movie;
eventually all the movies used were action, adventure, and science
fiction films.

Encouraged by the enthusiasm of the students, the authors decided
to take the project to the next level by considering possible
extensions of the course that would accommodate the curiosity of
every student and would satisfy the needs of every instructor.
Thus, we decided to create versions (packages) ---nicknamed
\textit{flavors}---of the course whereby each flavor used a
particular genre or theme of movies. As a result, plans were
developed to create the following flavors, each using films with a
well defined theme:

\begin{figure}[ht!]
\begin{center}
\includegraphics[width=3in]{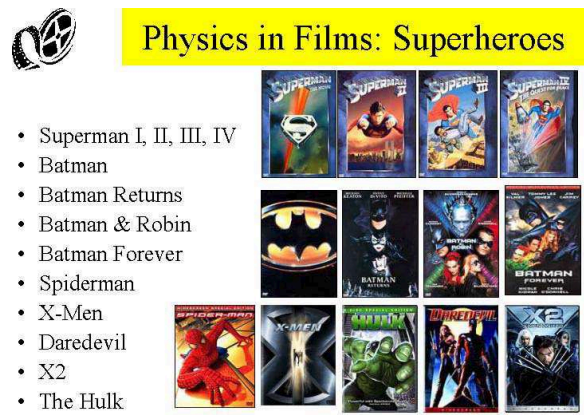}
\end{center}
\caption{Superheroes films used in the superhero
         approach of the \textit{Physics In Films}
         variation of the \textit{Physical Science} course.}
\label{fig5}
\end{figure}


\begin{enumerate}
\item \textbf{Action/Adventure} that would use action and adventure movies;
\item \textbf{Sci-Fi} that would use science fiction movies;
\item \textbf{Superhero} that would use superhero movies;
\item \textbf{Modern Physics} that would use movies that enable the teaching of topics from Modern Physics;
\item \textbf{Astronomy} that would use movies that contain topics related to astronomy;
\item \textbf{Pseudoscience} that would use movies including topics from pseudoscientific;
\item \textbf{Metaphysics} that would use movies that touch on questions of metaphysical content.
\end{enumerate}
The instructors started building the \textit{Superhero} and \textit{Pseudoscience} flavors in Summer 2003 and the
coming summer will finish their development. In figures 5 and 6, the reader
sees the movies used so far.

\begin{center}
\textbf{Example--}\textbf{\textit{Superman II}}
\end{center}

In \textit{Superman, the movie} the audience learns that before the planet Krypton exploded, three
criminals were sealed in a container and sent to the Phantom Zone (an
extraordinary prison) for eternity. All of them (on Earth) would possess
incredible powers ¯exactly equal to those of Superman. In \textit{Superman II,} extraordinary
conditions, of course, allow them to escape and arrive at Earth where they
terrorize the humans. During their trip towards Earth, they stop on the
Moon. There they become aware of their incredible powers and we witness a
discussion among them on their newly found powers. As real as this scene
seems to the audience, it could never have taken place. Human voice is a
sound wave that is created by vibrations of the vocal cords generating
density variations in the air inside the larynx. Sound waves can be created
only inside materials (such as the air of Earth's atmosphere) since they are
really changes in the density of the material. The Moon, however, has no
atmosphere; there is no material whose density can be affected to create
sound waves. Therefore, sound on the surface of the Moon (without the use of
sophisticated electronic equipment) is impossible. As has humorously been
said, ``in the vacuum of space, no one can hear you scream''.


\begin{figure}[h!]
\begin{center}
\includegraphics[width=3in]{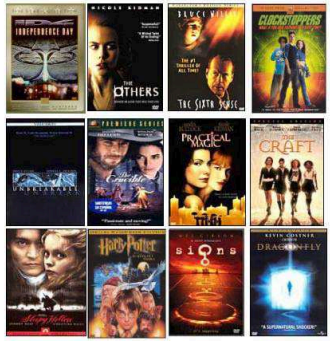}
\end{center}
\caption{Films with pseudoscientific content used in the
         pseudoscience approach of the \textit{Physics In
         Films}  variation of the \textit{Physical
         Science} course.}
\label{fig6}
\end{figure}


\begin{center}
\textbf{Example---}\textbf{\textit{Sixth Sense}}
\end{center}

\textit{The Sixth Sense} is a film concerned with ghosts. A child has the ability to see and
communicate with ghosts. The movie consistently tells the viewer that ghosts
like low temperatures, although why that should be is not explained. In a
scene where the young hero goes to the bathroom during the night, the
director clearly and distinctly shows a sudden drop in the room temperature,
so that one expects the appearance of a ghost; and indeed one appears.

To get a hint concerning the possibility that the appearance of
ghosts is heralded by a sudden drop in temperature, we shall look
at a case studied by scientists (BBC News 2001, Frood 2003). In
Hampton Court Palace near London, UK, there is a well-known
Haunted Gallery. It is said that The Gallery is stalked by the
spirit of Catherine Howard. Many visitors to the room have
described strange phenomena in the gallery such as hearing screams
and seeing apparitions. Due to many reports of such occurrences, a
team of `ghost-busting' psychologists, led by Dr Richard Wiseman
of Hertfordshire University was called to investigate. The team
installed thermal cameras and air movement detectors in the
gallery, following which about 400 palace visitors were asked if
they could feel a ``presence" in the gallery. The response was
quite amazing: more than half reported sudden drops in temperature
and some said they sensed a ghostly presence. Several people
claimed to have seen Elizabethan figures. However, the team
discovered that the experiences could be simply explained by the
gallery's numerous concealed old doors. These exits are far from
draught-proof and the combination of air currents which they admit
cause sudden changes in the room's temperature. In two particular
spots, the temperature of the gallery plummeted down to 36$^{o}$
F. ``You do, literally, walk into a column of cold air sometimes,"
said Dr Wiseman. Convection is one of the three ways heat
propagates; the other two are conduction and radiation. Convection
appears in fluids that have a non-uniform distribution of
temperature. As a result, currents inside the fluid will be such
as to attempt to restore a uniform temperature. These currents are
stronger when the non-uniformity is greater. In the case of the
gallery rooms, the convection currents would be felt as cold
drafts, similar to those experienced by someone who opens the door
of a refrigerator a hot day in summer.


\begin{figure}[tb!]
\begin{center}
\includegraphics[width=3in]{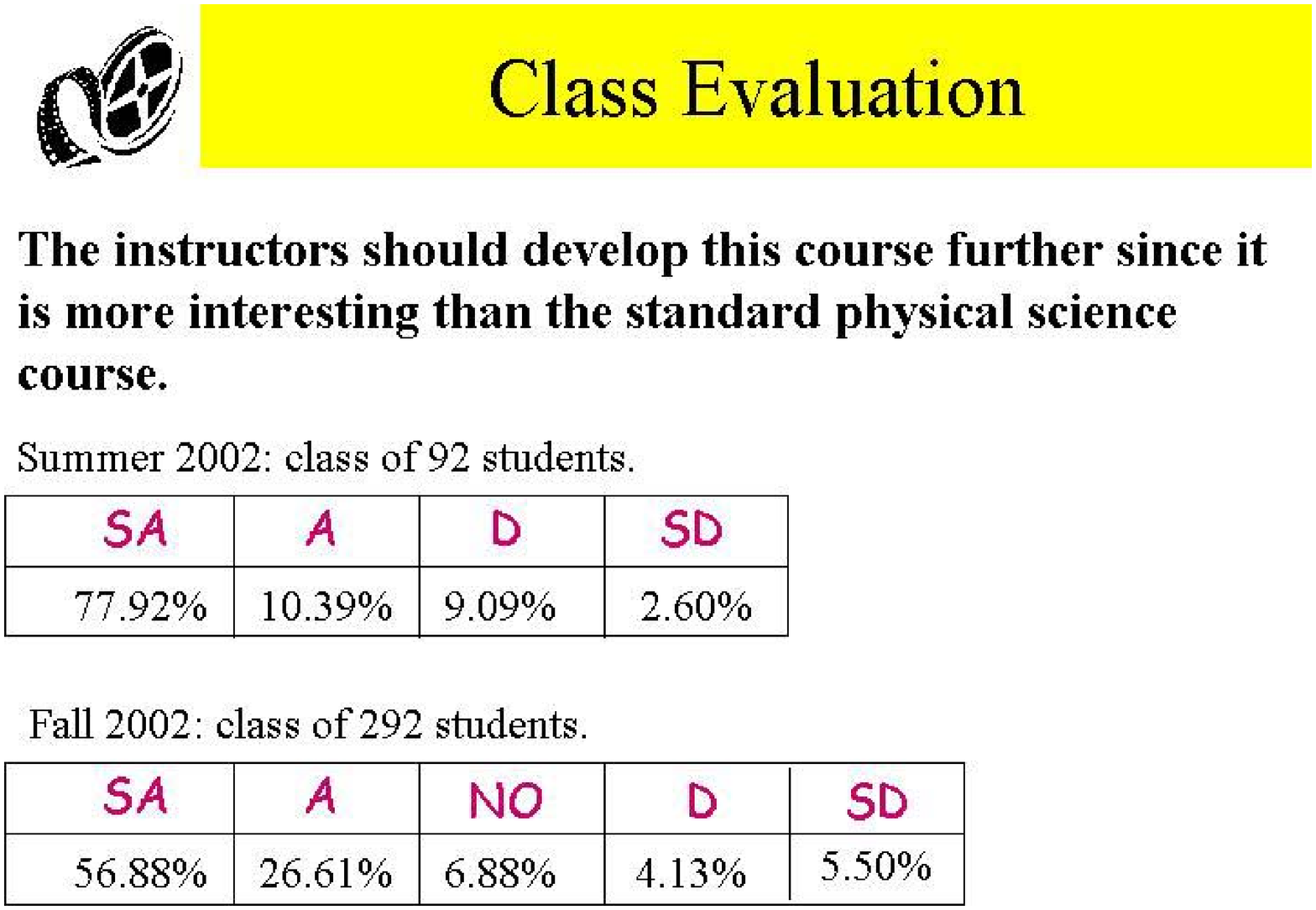}
\end{center}
\caption{One of the many class evaluation questions given to
         students. Notice that students strongly support the further
         development of the course. (SA=strongly approve, A=approve, NO=no
         opinion, D=disapprove, SD=strongly disapprove.) Statistics to
         other questions is similar. }
\label{fig7}
\end{figure}


\begin{center}
\textbf{THE RESULTS}
\end{center}

Student interest and performance in the \textit{Physical Science}
course have both increased dramatically compared with the
traditional teaching mode, which we still use in some sections. In
figures 7 and 8 we give an example of the several measures of
student interest that we have used, including student opinions of
the electronic personal response system used in class to answer
quiz questions and to take attendance (both of which contribute to
the final grade).

The exam scores distribution in two classes of about the same size (295
students each) taught by the same instructor, one in traditional lecture
mode, the other \textit{Physics in Films} mode and covering the same topics are quite dramatically
different, as the table below illustrates.


\begin{figure}[h!]
\begin{center}
\includegraphics[width=3in]{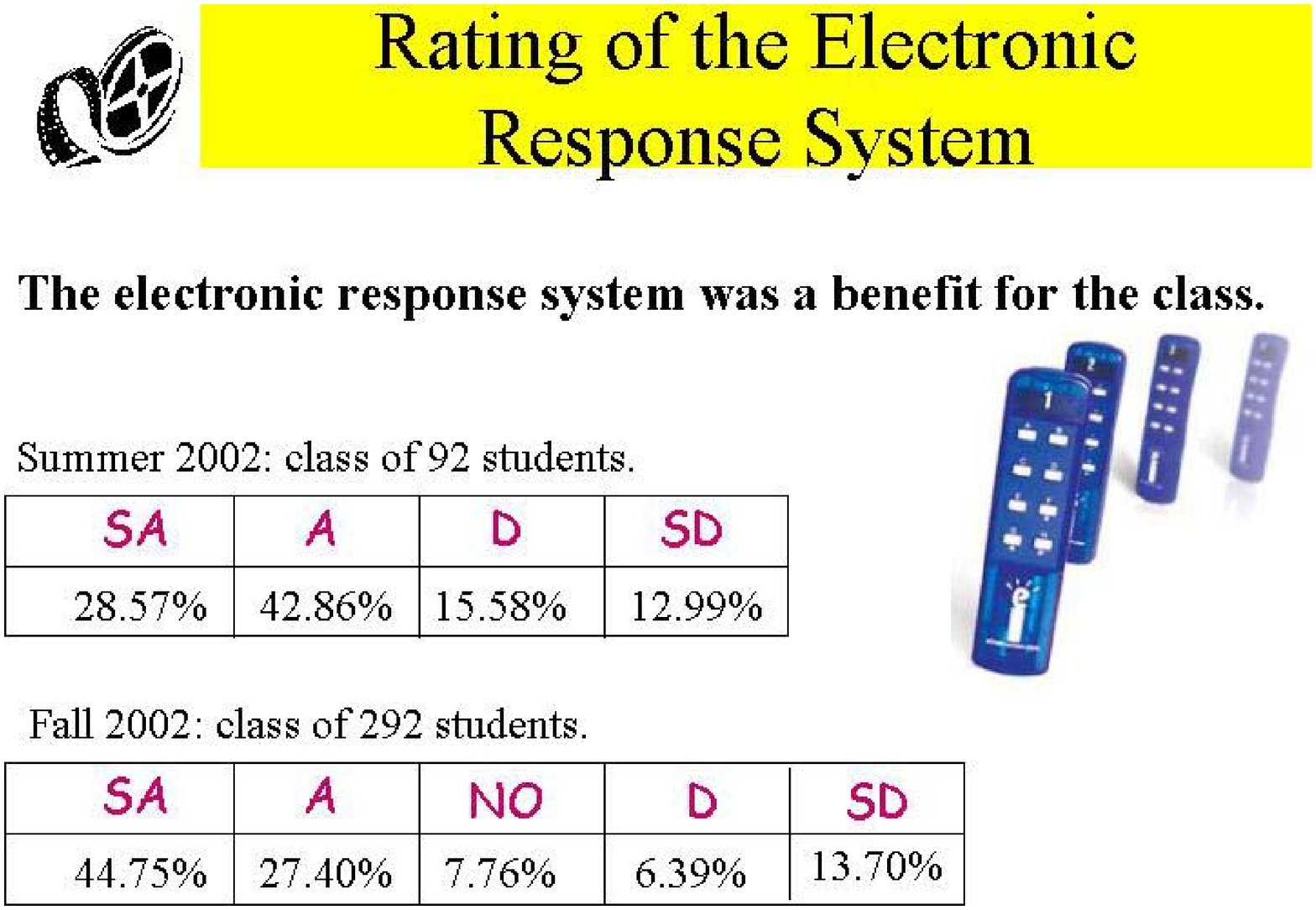}
\end{center}
\caption{The class is using a personal response system in order to
         obtain real-time data on the class state of knowledge. Although
         the students at the beginning of the class complain about the use
         of the system (since it increases their expenses for the course
         and forces them to attend the lectures), by the end of the course
         are very enthusiastic about it.}
\label{fig8}
\end{figure}

\begin{table}
\begin{center}
\begin{tabular}{|c|c|c|c|}\hline
\multicolumn{4}{|c|}{\bfseries Fall 2002 Traditional mode}\\
                                                             \hline
        Exam 1 & Exam 2 & Exam 3 & Final \\ \hline
 49.3   & 65.3   & 58.2   & 59.4  \\ \hline
\end{tabular}

\vspace{5mm}
\begin{tabular}{|c|c|c|c|}\hline
\multicolumn{4}{|c|}{\bfseries Fall 2003 `Physics in Films' mode}\\
                                                             \hline
        Exam 1 & Exam 2 & Exam 3 & Final \\ \hline
  74.9   & 67.7   & 75.7   & 72.8  \\ \hline
\end{tabular}
\end{center}
\caption{Exam averages (100 being the maximum score)of two large
         sections of \textit{Physical
         Science} taught by the same
         instructor.}
\end{table}


\begin{center}
\textbf{THE FUTURE}
\end{center}

Our goal is to increase awareness of science and to demonstrate
that understanding some physics can be exciting and rewarding when
presented in the context of an enjoyable activity. Our
presentation has illustrated in detail exactly how the movies are
used as a vehicle for learning physical science. In class clips
from the films are integrated with lecture materials, quizzes,
demonstrations, and slides so as to enhance both student attention
and retention. We have given two sample scene analyses and
comparative statistics of student reactions and performance. This
new technique can be extended to many other disciplines, as the
following list  (with some suggested films) illustrates.
\begin{enumerate}
 \item \textbf{\textit{Math in Films}}: Pi, Good Will Hunting, Pay
           it Forward, A Beautiful Mind
 \item \textbf{\textit{Biology in Films}}:
           Spiderman, Planet of the Apes, Fly, Hollow Man
 \item \textbf{\textit{Engineering in Films}}:
            Armageddon, The Bridge on River Kwai, Space Cowboys
 \item \textbf{\textit{Archeology in Films}}: Indiana Jones: Raiders of the
            Lost Ark, The Mummy, The Relic
 \item \textbf{\textit{Computers in Films}}:
            Independence Day, The Net, Swordfish
 \item \textbf{\textit{Philosophy in Films}}: Matrix, Terminator, Blade
            Runner
 \item \textbf{\textit{History in Films}}: Patriot, Braveheart, Gladiator,
            Elizabeth
 \item \textbf{\textit{Law in Films}}: Legally Blonde, Erin
            Brockovich, The Firm, The Rainmaker
 \item \textbf{\textit{Forensic Science in Films}}: Murder by the Numbers, Bone
            Collector, Torso, Jennifer 8
 \item \textbf{\textit{Psychology in Movies}}: Don't Say a Word, Control,
            Final Analysis, Primal Fear
\end{enumerate}
The authors are in the process of preparing a self-contained physical
science course, complete with textbook and a CD-ROM with all scene analyses
and slides we use, that can be handed to an instructor ready-to-go.



\end{document}